\documentclass{IEEEtran4PSCC}
\ifCLASSINFOpdf
   \usepackage[pdftex]{graphicx}
\else
   \usepackage[dvips]{graphicx}
\fi

\usepackage[cmex10]{amsmath}
\usepackage{amsfonts}
\usepackage{amssymb}
\usepackage{tensor}
\usepackage{bbm}
\usepackage{lipsum} 
\usepackage{xcolor}
\usepackage{booktabs}  
\usepackage{algorithm}
\usepackage{algpseudocode}
\usepackage{bbold}  
\usepackage{cite}

\usepackage{todonotes}
\usepackage{mathtools}
\usepackage{etoolbox}
\makeatletter

\patchcmd{\@IEEEeqnarray}{\relax}{\relax\intertext@}{}{}
\makeatother

\usepackage{glossaries-extra}
\setabbreviationstyle[acronym]{long-short}  
\newacronym{fp}{FPI}{fixed-point iteration}
\newacronym{pf}{PF}{power flow}
\newacronym{tpf}{TPF}{TensorPowerFlow}
\newacronym{nr}{NR}{Newton-Raphson}
\newacronym{cpu}{CPU}{central processing unit}
\newacronym{gpu}{GPU}{graphics processing unit}
\newacronym{ppf}{PPF}{probabilistic power flow}
\newacronym{sam}{SAM}{successive approximation method}
\newacronym{blas}{BLAS}{basic linear algebra subprograms}
\newacronym{tss}{TSS}{time series simulation}

\usepackage{url}

\usepackage{graphicx} 
\usepackage{import}
\usepackage{xifthen}
\usepackage{pdfpages}
\usepackage{transparent}
\graphicspath{{Figures}}  

\makeatletter
\let\old@ps@headings\ps@headings
\let\old@ps@IEEEtitlepagestyle\ps@IEEEtitlepagestyle
\def\psccfooter#1{%
    \def\ps@headings{%
        \old@ps@headings%
        \def\@oddfoot{\strut\hfill#1\hfill\strut}%
        \def\@evenfoot{\strut\hfill#1\hfill\strut}%
    }%
    \def\ps@IEEEtitlepagestyle{%
        \old@ps@IEEEtitlepagestyle%
        \def\@oddfoot{\strut\hfill#1\hfill\strut}%
        \def\@evenfoot{\strut\hfill#1\hfill\strut}%
    }%
    \ps@headings%
}
\makeatother

\begin{document}
\title{Tensor Power Flow Formulations for Multidimensional Analyses in Distribution Systems}

\author{
\IEEEauthorblockN{
Edgar Mauricio Salazar Duque\IEEEauthorrefmark{1},
Juan S. Giraldo\IEEEauthorrefmark{2}, 
Pedro P. Vergara\IEEEauthorrefmark
{3},
Phuong H. Nguyen\IEEEauthorrefmark{1},
Han (J.G.) Slootweg\IEEEauthorrefmark{1}
}

\IEEEauthorblockA{
\IEEEauthorrefmark{1} Eindhoven University of Technology, Netherlands,
\IEEEauthorrefmark{3} Delft University of Technology, Netherlands}
\IEEEauthorblockA{\IEEEauthorrefmark{2} Energy Transition Studies Group, Netherlands Organisation for Applied Scientific Research (TNO), Netherlands}
}

\maketitle

\begin{abstract}
In this paper, we present two multidimensional power flow formulations based on a fixed-point iteration (FPI) algorithm to efficiently solve hundreds of thousands of \glspl{pf} in distribution systems. The presented algorithms are the base for a new TensorPowerFlow (TPF) tool and shine for their simplicity, benefiting from multicore \gls{cpu} and \gls{gpu} parallelization. We also focus on the mathematical convergence properties of the algorithm, showing that its unique solution is at the practical operational point, which is the solution of high-voltage and low-current. The proof is validated using numerical simulations showing the robustness of the FPI algorithm compared to the classical \gls{nr} approach. In the case study, a benchmark with different PF solution methods is performed, showing that for applications requiring a yearly simulation at 1-minute resolution the computation time is decreased by a factor of 164, compared to the NR in its sparse formulation.
\end{abstract}



\section{Introduction}
The power flow study is crucial for different technical analyses of electrical distribution systems. For example, power flows are widely used in  \gls{tss}, where long-term analysis depends on high granularity in time, e.g., integration of distributed energy resources, Volt/Var control, and hosting capacity~\cite{Qureshi_2019}. Another widespread use of multiple power flows is the \gls{ppf}, in which exogenous uncertainties are modeled using scenarios, evaluated using power flows, and its impact evaluated using stochastic analysis~\cite{liu2018probabilistic}. Many other applications require multiple power flow executions, such as metaheuristic-based optimization, contingency analysis, and machine learning in power systems~\cite{duque2022community}.

Notably, the common factor between the aforementioned applications is the execution of multiple (thousands or even millions) power flows to provide insightful results. These applications are \textit{multidimensional} regardless of the technique used to reduce the number of timestamps, scenarios, or training size~\cite{Qureshi_2019, liu2018probabilistic, duque2022community}. The multidimensionality motivates efficient power flow formulations and new techniques, providing fast and accurate results. Multidimensional problems have been tackled in the past, such as in~\cite{liu2018probabilistic} for the \gls{ppf} problem using the embedded holomorphic power flow, or in~\cite{Garces_2019} where tensors are used to formulate the three-phase power flow. However, they are not scalable or even computationally more exhaustive than traditional alternatives. 

Advances in computer hardware, such as the increase in the number of cores in \glspl{cpu} and the evolution of \gls{gpu} designs, transitioning from simple graphics processors to highly parallel multiprocessors of many cores \cite{nickolls_gpu_2010}, opened a new paradigm of programming and rethinking PF algorithms. New lines of research look at reducing computational time by combining CPU and GPU resources to improve the speed of convergence of Newton-Raphson algorithms \cite{zhou_gpu_2017, li_gpu_2017, zhou_gpu_2018, zhou_gpu_acc_2017}, which is seen as a preferred method to improve due to its quadratic convergence. However, the formulation of these approaches is designed specifically to solve one power flow in the least amount of time. Unfortunately, these algorithms are heavily focused on GPU-based architectures, making their implementation a tedious process and requiring specific GPU programming knowledge.

In this paper, we advocate for using a fixed point iteration algorithm to solve multidimensional PF formulations. This fixed point algorithm has shown robust performance \cite{giraldo2022fixed}; it has a simple formulation for the case of distribution system analyses, it is suitable and scalable for multidimensional applications in the form of a tensor, and can also benefit from multicore CPU and GPU parallelization. In this work, we also focus on the mathematical convergence properties of the algorithm, showing its unique point of solution at the practical operational point (if the solution exists), which is the high-voltage, low-current solution. The contributions of the paper are as follows:
\begin{itemize}
    \item Present a practical tool for multidimensional power flow analysis in distribution systems, named \gls{tpf}\footnote{The public link to the repository will be provided after the acceptance of the paper} in its dense and sparse versions, which are based on a \gls{fp} algorithm. The \gls{tpf} opens new possibilities for using mixed computing resources (\glspl{cpu} or \glspl{gpu}) to increase performance. 
    \item Give a geometric and physical interpretation of the existence of the power flow solution using an FPI iterative algorithm, showing the robustness and numerical stability of the FPI algorithm.
\end{itemize}




\section{Single Dimension Fixed Point Power Flow}\label{sec:single_dim}
We take a network with one substation, $b=|\Omega_{\mathrm{d}}|$ demand nodes, and \mbox{$\phi=|\Omega_{\phi}|$} phases. Nodal voltages and currents are related by the admittance matrix $\boldsymbol{Y}\in\mathbb{C}^{(b+1)\,\phi\times (b+1)\,\phi}$ as follows:

\begin{flalign}
\label{eq:rect_coord}
\small
    \begin{bmatrix}  \boldsymbol{i}_{s}\\
                    -\boldsymbol{i}_{d}
    \end{bmatrix}
    =
    \begin{bmatrix} \boldsymbol{Y}_{ss} & \boldsymbol{Y}_{sd}\\
                    \boldsymbol{Y}_{ds} & \boldsymbol{Y}_{dd}
    \end{bmatrix}
    \begin{bmatrix} \boldsymbol{v}_{s}\\
                    \boldsymbol{v}_{d}
    \end{bmatrix}
\end{flalign}

\noindent where vectors ${{\boldsymbol{i}}_{{s}}}\in\mathbb{C}^{\phi\times 1}$ and ${{\boldsymbol{i}}_{{d}}}\in\mathbb{C}^{b\phi\times 1}$ represent complex injections of nodal current at the substation and at the demand nodes $d\in \Omega_{\mathrm{d}}$, while ${{\boldsymbol{v}}_{{s}}}\in\mathbb{C}^{\phi\times 1}$ and ${\boldsymbol{v}}:={{\boldsymbol{v}}_{{d}}\in\mathbb{C}^{b\phi\times 1}}$ are complex components of the respective nodal voltages. It must be noted that~\eqref{eq:rect_coord} is general and can represent single-phase, polyphase, radial, meshed networks and distributed generation of constant power. 

In this paper, we call a \textit{single-dimensional} power flow to the solution of \eqref{eq:rect_coord} for a single vector of nominal complex power $\boldsymbol{s}\in\mathbb{C}^{b\phi\times 1}$ at the demand nodes, which is a snapshot of consumption in the grid. The single-dimensional power flow formulation, namely the \gls{fp} algorithm, is the basis of the tensorized version and it is based on the notation of the equivalent version of the \gls{sam} presented in~\cite{giraldo2022fixed} as:
\begin{flalign}
\label{eq:fp_algorithm}
\boldsymbol{v}_{(n+1)} = {\boldsymbol{F}\boldsymbol{v}_{(n)}^{*}}^{\circ(-1)} + \boldsymbol{w}
\end{flalign}
\noindent with
\begin{flalign}
&\boldsymbol{A} = \textbf{\textrm{diag}}\left(\alpha_{P}\!\odot\!\boldsymbol{s}^{*}\right);\hspace{15pt}  \boldsymbol{B} = \textbf{\textrm{diag}}\left({\alpha_{Z}\odot\boldsymbol{s}^{*}}\right)+ \boldsymbol{Y}_{dd} ;\nonumber\\
\label{eq_b}
&\boldsymbol{c} = \boldsymbol{Y}_{ds}\boldsymbol{v}_{s} +  {\alpha_{I}\odot\boldsymbol{s}^{*}}; \nonumber \\
&\boldsymbol{F} = -\boldsymbol{B}^{-1}\boldsymbol{A}; \hspace{20pt} \boldsymbol{w}=-{\boldsymbol{B}}^{-1}\boldsymbol{c} 
\end{flalign}

\noindent where $\odot$ is the Hadamard product, ${\boldsymbol{v}_{(n)}^{*}}^{\circ(-1)}$ the element-wise reciprocal of the conjugated vector containing the nodal voltages at iteration $n$, and $\alpha_Z,\,\alpha_I$, and $\alpha_P$ represent the coefficients of the ZIP load model per demand node and phase. It is worth mentioning that $\boldsymbol{A}\in\mathbb{C}^{b\phi\times b\phi}$,  $\boldsymbol{B}\in\mathbb{C}^{b\phi\times b\phi}$, and \mbox{$\boldsymbol{c}\in\mathbb{C}^{b\phi\times 1}$} are constant matrices and vector within the iterative process for a particular operating point. Thus, $\boldsymbol{F} \in \mathbb{C}^{b\phi\times b\phi}$ and $\boldsymbol{w} \in \mathbb{C}^{b\phi\times 1}$ can be calculated once used at each iteration. 

The algorithm in \eqref{eq:fp_algorithm} can be rearranged as an iterative mapping using $\boldsymbol{v}_{(n)}^{*\circ{-1}}=\boldsymbol{v}_{(n)} \oslash \|\boldsymbol{v}_{(n)}\|^{2}_{\circ}$, where $\oslash$ is the Hadamard division and $\|\cdot\|_{\circ}$ is the element-wise euclidean norm, as
\begin{IEEEeqnarray}{rCl}
     \boldsymbol{v}_{(n+1)} &=& -\boldsymbol{B}^{-1}\;\textbf{\textrm{diag}}(\alpha_p \odot \boldsymbol{s}^* \oslash \|\boldsymbol{v}_{(n)}\|^{2}_{\circ}) \boldsymbol{v}_{(n)} + \boldsymbol{w} \nonumber \\
                            &=& T(\boldsymbol{v}_{(n)}) \label{eq:contraction_mapping}
\end{IEEEeqnarray}

Here, the Banach fixed-point theorem 
is used to prove that~\eqref{eq:contraction_mapping} is a contraction mapping in order to have a solution. This is the case where $T(\cdot)$ satisfies $\|\boldsymbol{v}_{(n+1)} - \boldsymbol{v}_{(n)}\|_1 \leq k\|T(\boldsymbol{v}_{(n+1)})-T(\boldsymbol{v}_{(n)})\|_1$, where the norm-1 distance is used in our case ($\|\cdot\|_1$), that is, for vectors $\|\boldsymbol{x}\|_1 = \sum_{i} |x_i|$, where $|\cdot|$ is the absolute value, and for matrices $\|A\|_1 = \max_j(\sum_i |a_{ij}|)$, which is the column norm. When $k < 1$, the contraction mapping $T(\cdot)$ has a unique point of convergence. This can be shown as
%
\begin{IEEEeqnarray}{rCl}
    \IEEEeqnarraymulticol{3}{l}{ \|\boldsymbol{v}_{(n+1)} - \boldsymbol{v}_{(n)} \|_1} \nonumber\\
    \quad &=& \|\boldsymbol{B}^{-1}\;\textbf{\textrm{diag}}(\alpha_p \odot \boldsymbol{s}^* \oslash \|\boldsymbol{v}_{(n+1)}\|^{2}_{\circ}) \boldsymbol{v}_{(n+1)} \nonumber \\
       && -\>\boldsymbol{B}^{-1}\;\textbf{\textrm{diag}}(\alpha_p \odot \boldsymbol{s}^* \oslash \|\boldsymbol{v}_{(n)}\|^{2}_{\circ}) \boldsymbol{v}_{(n)}\|_1 \label{eq:banach}
\end{IEEEeqnarray}
\noindent where the conjugate load power in each bus-phases is expressed by its equivalent load impedance solution $\boldsymbol{z}_l$. i.e., $\boldsymbol{s}^* \oslash \|\boldsymbol{v}_{(n)}\|^2_{\odot}= \mathbbm{1} \oslash \boldsymbol{z}_{l}$, where $\mathbbm{1}$ is a vector of ones of dimension $b\phi$. Assuming that $\boldsymbol{s}$ is a feasible load consumption in the grid (details of this feasibility are given in Section~\ref{sec:geometry}), the iterative algorithm reaches closer voltage values for each iteration, i.e., $\boldsymbol{v}_{(n+1)} \mapsto \boldsymbol{v}_{(n)}$, then \eqref{eq:banach} can be reduced using Hölder's inequality as 
\begin{IEEEeqnarray}{rCl}
    \|\boldsymbol{v}_{(n+1)} - \boldsymbol{v}_{(n)} \|_1 &=& \|\boldsymbol{B}^{-1}\;\textbf{\textrm{diag}}(\alpha_p \oslash \boldsymbol{z}_{l}) (\boldsymbol{v}_{(n+1)}-\boldsymbol{v}_{(n)}) \|_1 \nonumber\\
     &\leq& \|\boldsymbol{B}^{-1}\;\textbf{\textrm{diag}}(\alpha_p \oslash \boldsymbol{z}_{l}) \|_1 \| (\boldsymbol{v}_{(n+1)}-\boldsymbol{v}_{(n)}) \|_1 \nonumber \\
     &\leq& k \| (\boldsymbol{v}_{(n+1)}-\boldsymbol{v}_{(n)}) \|_1 \label{eq:contraction}
\end{IEEEeqnarray}
From \eqref{eq:contraction}, for purely constant power $\alpha_p=1$, and noticing that $\boldsymbol{B}^{-1}$ is the grid impedance matrix $\boldsymbol{Z}_{\mathrm{B}}$. Then, the contraction scalar is
\begin{IEEEeqnarray}{rCl}
    k&=& \|\boldsymbol{Z}_{\mathrm{B}}\textbf{\textrm{diag}}(\mathbbm{1}\oslash\boldsymbol{z}_{l})\|_1  = |\hat{z}_{jj}|/|\hat{z}_{l,j}| \label{eq:ratio}
\end{IEEEeqnarray}
where the hat notation represents the solution values of the biggest ratio between impedances. The only possible solution of \eqref{eq:ratio} in order to be a contraction mapping, i.e., $k<1$, is when $|z_{jj}| < |z_{l,j}|$. The diagonal entries of $\boldsymbol{Z}_{\mathrm{B}}$, i.e., $|z_{jj}|$, are the Thevenin impedance equivalent of bus $j$ \cite{grainger_power_1994}, meaning that the only solution for \eqref{eq:ratio} is for the operational point of the network with high impedance (high voltage, low current), which is the feasible operational state of the network. The physical interpretation of \eqref{eq:ratio} can be shown by analysis of an equivalent two-bus system, which is discussed in detail in the next section.

\begin{figure*}[t]
\centering
\includegraphics[]{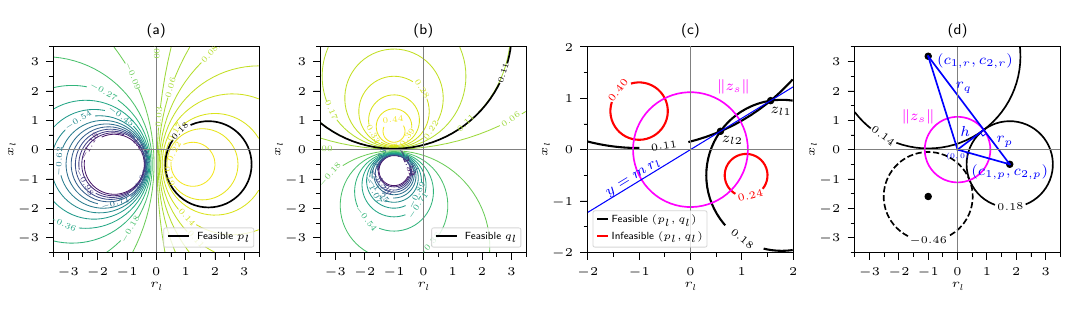}
\caption{Geometry of the solutions of the power flow problem for a two-bus system with parameters $z_s=1.0 + \boldsymbol{j}0.5$ and $\|v_o\|=1.0$. (a) and (b) are the contour graphs for the circular formulation~\eqref{eq:circle_pl} and~\eqref{eq:circle_ql}, for $p_l$ and $q_l$, respectively; black circles highlight a feasible combination of active and reactive power. (c) The red circles highlight an infeasible value of the load power (no crossing between the circles), while the feasible values have two points of solution that form a line that passes through the origin. (d) All the solutions for the critical load power, which has only one point of contact between the circles, form a circle with radius $\|z_s\|$.}
\label{fig:geometry}
\end{figure*}

\section{Geometric interpretation of the existence of a power flow solution}\label{sec:geometry}
\begin{figure}[t]
\centering
\includegraphics[width=0.85\columnwidth]{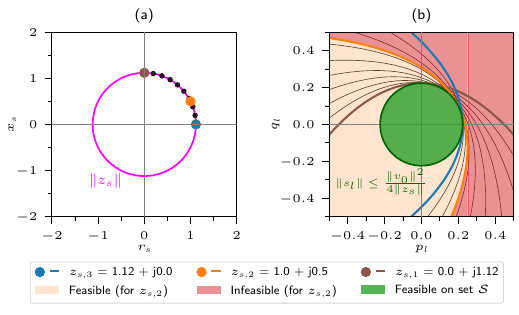}
\caption{Example of feasibility regions depicted by \eqref{eq:conic_section} and \eqref{eq:region}. (a) Each point of the maximum power transfer circle $\|z_s\|$ in the impedance plane $(r_s, x_s)$ parametrize the parabola in \eqref{eq:conic_section} that defines the feasible load power values.  (b) The vertexes of the parametrized parabolas are rotating around the circle defined by \eqref{eq:region} in the load power plane $(p_l,q_l)$. Only points of the first quadrant are shown, highlighting three example points ($z_{s,1}$, $z_{s,2}$, $z_{s,3}$) with their respective parabolas. Feasibility region for the case of $z_{s,2}$ is emphasised with red and yellow. The green region is the union of all possible feasible regions for all parabolas.}
\label{fig:parabolas}
\end{figure}

Consider a grid formed by two nodes: node 0 as a source node, acting as the slack bus with known voltage $v_0$ and reference angle $\theta_0=0$, and node 1 with load $s_l$, operating as a load bus. Then, the power of the load is given by $s_l=v_l i_l^*$, defining $v_l=(z_l/(z_s + z_l))v_0$, $i_l=(v_0/(z_s+z_l))$, where the source load is $z_s=r_s + \boldsymbol{j}x_s$, and the equivalent load impedance is $z_l = r_l + \boldsymbol{j}x_l$. The power on the load defined in terms of the source voltage and impedances is
\begin{IEEEeqnarray}{rCl}
    s_l &=& p_l + \boldsymbol{j}q_l = \big(\frac{r_l}{a^2+b^2}+\boldsymbol{j}\frac{x_l}{a^2+b^2}\big)\|v_0\|^2,
\end{IEEEeqnarray}
where $a=(r_s+r_l)$ and $b=(x_s+x_l)$. Rearranging the expressions for $p_l$ and $q_l$ into circle equations as \eqref{eq:circle_pl} and \eqref{eq:circle_ql} respectively, we have that
\begin{IEEEeqnarray}{rCr}
    (r_l - c_{1,p})^2 + (x_l - c_{2,p})^2 &=& r_p^2 \label{eq:circle_pl}\\
    (r_l - c_{1,q})^2 + (x_l - c_{2,q})^2 &=& r_q^2 \label{eq:circle_ql}
\end{IEEEeqnarray}
\begin{align*}
    c_{1,p} &= \frac{\|v_0\|^2}{2p_l}-r_s &  c_{1,q} &= -r_s \\
    c_{2,p} &= -x_s                     &  c_{2,q} &= \frac{\|v_0\|^2}{2q_l} - x_s \\
    r_p     &= \frac{\|v_0\|}{2p_l}\sqrt{\|v_o\|^2-4r_sp_l}  & r_q &= \frac{\|v_0\|}{2q_l}\sqrt{\|v_o\|^2-4x_sq_l}
\end{align*}
The contour plots of \eqref{eq:circle_pl} and \eqref{eq:circle_ql} are shown in Fig.~\ref{fig:geometry}(a,b), highlighting as an example the circles for the power values $p_l=0.18$ [p.u] and $q_l=0.11$ [p.u]. The intersections of the circles are the two possible impedance solutions for $z_l$, i.e., $z_{l,1}$ and $z_{l,2}$, where the values for $p_l$ and $q_l$ can exist simultaneously. For example, Fig.~\ref{fig:parabolas}(c) shows two cases: the first is where the power flow has a solution and the circles intersect (black line) at two points, and the second is where the red circle does not have an intersection, meaning an infeasible combination of power values. The intersection points of the circles form a line that passes through the origin in all cases. This can be shown using the intercept in the line equation $y=mr_l+\beta$, shown as a blue line in Fig.~\ref{fig:geometry}(c). The intercept is defined in terms of the parameters of the two circles as
\begin{IEEEeqnarray}{rCl}\label{eq:intersection}
    \beta &=& \frac{\overbrace{(c_{1,p}^2 + c_{2,p}^2 + r_q^2)}^{B_0} - \overbrace{(c_{1,q}^2+c_{2,q}^2 + r_p^2)}^{B_1}}{2(c_{2,p} - c_{2,q})}
\end{IEEEeqnarray}
where the numerator of \eqref{eq:intersection} is zero because $B_0=B_1$.

The worst-case loading scenario occurs when the two circles have only one point of contact (Fig.~\ref{fig:geometry}(d)). The tangent line of the point of contact between the circles is perpendicular to their radii and also passes through the origin according to \eqref{eq:intersection}. Therefore, the altitude ($h$) of the scalene triangle formed by the centers of the circles and the origin, that is, points $\langle (c_{1,p}, c_{2,p}), (c_{1,q}, c_{2,q}), (0,0) \rangle$, can be calculated using the triangle inequalities as
\begin{IEEEeqnarray}{rCr}
    h^2 &=& (c_{1,p})^2 + (c_{2,p})^2 -r_{p}^2 = r_s^2 + x_s^2 = \|z_s\|^2.
\end{IEEEeqnarray}

This means that all unique points of contact between the two circles lie in another circle defined by $\|z_s\|$, which is the magenta circle shown in Fig.~\ref{fig:geometry}(c,d). This also means that the critical points where the power flow is feasible and with only one impedance solution $z_l$, $z_{l,1}=z_{l,2}$, are when $\|z_s\|=\|z_l\|$, which is the point of maximum power transfer. From this, it is clear and \textit{important to notice} that there are two possible solutions for the two-bus system that are not critical; one lives inside and the other outside the circle defined by $\|z_s\|$, meaning that $\|z_{l,1}\|<\|z_s\|$ and $\|z_{l,2}\|>\|z_s\|$.

A region of convergence in the complex power plane $(p_l, q_l)$ can be defined for the power in the load $s_l$ using the critical point of contact of the two circles in the impedance plane $(r_l, x_l)$. When the circles have one point of contact, then the distance between the centers of the circles equals the sum of its radii, that is, $r_p + r_q=\sqrt{(c_{1,p} - c_{1,q})^2 + (c_{2,p} - c_{2,q})^2}$. Rearranging this expression we have the following quadratic equation, named $f_{\alpha}(p_l, q_l)|_{(r_s,x_s)}$, for $p_l$ and $q_l$
\begin{IEEEeqnarray}{rCr}
     G q_l^2 + Hp_lq_l + I p_l^2 + J q_l + K p_l +  L &=&0 \label{eq:conic_section}
\end{IEEEeqnarray}
\begin{align*}
    G &= r_s^2     & H &= -2r_sx_s       &  I &= \|v_0\|^2 x_s\\
    J &= x_s^2     &  K &= \|v_0\|^2 r_s &  L &= -\|v_0\|^2 / 4\\
\end{align*}

The conic section defined by \eqref{eq:conic_section} is a parabola due to its discriminant $\Delta = H^2-4GI=0$. Parabola is parametrized by the source impedance $z_s$, i.e., $r_s$ and $x_s$. An example of this parabola where $z_s$ has only a resistive value is shown in blue in Fig.~\ref{fig:parabolas}(a). The parabola curve represents all $(p_l, q_l)$ values where the two-bus system has the maximum power transfer, which means that it has only one equivalent load impedance solution. The region defined on the left side of the parabola is all the power values that have a valid solution, meaning that $z_{l,1}$ and $z_{l,2}$ exist, and the right regions are values for which the system is infeasible. This region changes as a result of the parameterization values of the parabola. The blue parabola is when $z_l$ has only a reactance value. If we compute all sets of parabolas $\mathcal{S}=\{f_{\alpha}(p_l, q_l)|_{(r_s, x_s)} \: | \: r_s^2 + x_s^2 = \|z_s\|^2\}$ and have the union of all feasible power flow regions defined by $\mathcal{S}$ we have the circle delimiting such a region by $\|s_l\| \leq \frac{\|v_0\|^2}{4\|z_s\|}$. This circle is the collection of all the vertices of the rotating parabolas. This means that the condition to have a feasible solution for the power flow is
\begin{IEEEeqnarray}{rCr}
    \|v_0\|^2 &\geq& 4\|s_l\|\|z_s\|. \label{eq:region}
\end{IEEEeqnarray}
This condition that is in the norm-form, is used to prove the existence of the power flow solution in \cite{sur_existence_2019,bolognani_existence_2016, yu_simple_2015}, and here it shows a geometrical derivation and interpretation. It should be noted that this region \eqref{eq:region} is a conservative estimate, since the actual regions depend on specific parameterization of $r_s$ and $x_s$~\cite{yu_simple_2015}, as shown in Fig.~\ref{fig:parabolas}(b) for the $z_{s,2}$ example. 

Therefore, if the power consumption in the grid follows at least the condition \eqref{eq:region}\cite{bolognani_existence_2016}, and the power flow converges using the iterative fixed point algorithm in \eqref{eq:fp_algorithm}, and following that the condition \eqref{eq:ratio} must be less than one, this implies that the unique point of convergence is the high-impedance solution.

\begin{figure}[t]
\centering
\includegraphics[width=0.9\columnwidth]{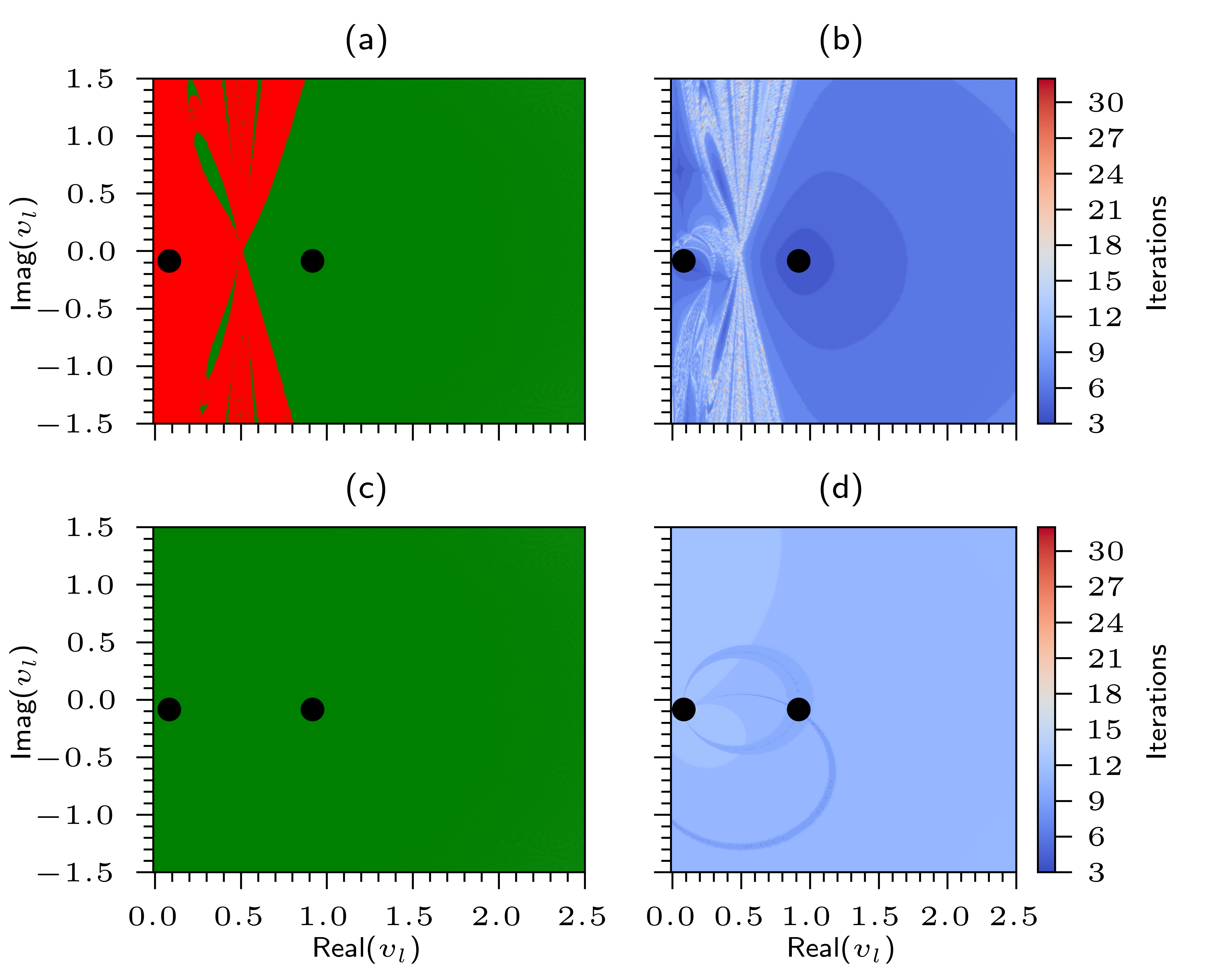}
\caption{Convergence analysis of NR and FPI algorithms. (a) and (c) corresponds to the regions of convergence for different initial values of voltages $v_0$ for NR and FPI algorithms, respectively. The green region corresponds to the high voltage (high impedance) and the red region to the low voltage (low impedance) solutions. (b) and (d) correspond to the number of iterations required for convergence, for the NR and FPI algorithms, respectively.}
\vspace{-3mm}
\label{fig:convergence}
\end{figure}

The two-bus system is solved using the \gls{nr} and \gls{fp} algorithms, with different starting points $\boldsymbol{v}_{0}$ to numerically confirm the unique convergence point of \eqref{eq:fp_algorithm}. The first row of subplots in Fig.~\ref{fig:convergence} corresponds to the NR and the second \gls{fp} solutions. The green and red colors in Fig.~\ref{fig:convergence}(a,c) represent the attraction regions for high-voltage (high-impedance) and low-voltage solutions, respectively. The number of iterations for each algorithm is shown as colored contour plots in Fig.~\ref{fig:convergence}(b,d) for the \gls{nr} and \gls{fp} algorithms, respectively. The results confirm the robustness of the FPI algorithm, which converges to the operational point irrelevant from the initial voltage estimate, unlike \gls{nr}, on which for a simple two-bus system, the risk of falling into a nonoperational point (but mathematically valid) still exists and is dependent on the $\boldsymbol{v}_0$ starting point. Additionally, the \gls{fp} algorithm has a consistent number of iterations in the entire complex domain for $v_l$. The interested reader is referred to~\cite{giraldo2022fixed} for further numerical comparisons. 


The \gls{nr} algorithm in polar coordinates needs fewer iterations than the \gls{fp} approach \cite{giraldo2022fixed, ahmadi_fast_2022}. Moreover, \gls{nr} benefits from system sparsity, making it ideal for sequential computers. Still, its per-iteration computation cost is significantly higher due to the need for Jacobian inverse calculations. When dealing with hundreds to millions of power flows, the cumulative cost of \gls{nr} iterations adds up, leading to longer processing times. Thanks to advancements in computer hardware, especially faster matrix multiplications, the \gls{fp} algorithm is an excellent choice for extensive power flow simulations on modern computers for distribution networks. Its reliance on successive matrix multiplications, along with its simple formulation, makes it easy to program. Moreover, it can naturally extend to a tensor setting.

\section{Multidimensional Fixed Point Power Flow 
}\label{sec:multi_dim}
\subsection{Tensor Power Flow - Dense Formulation}
\begin{figure}[t]
    \centering
    \fontsize{5pt}{5pt}\selectfont
    \def\svgwidth{0.90\columnwidth}
    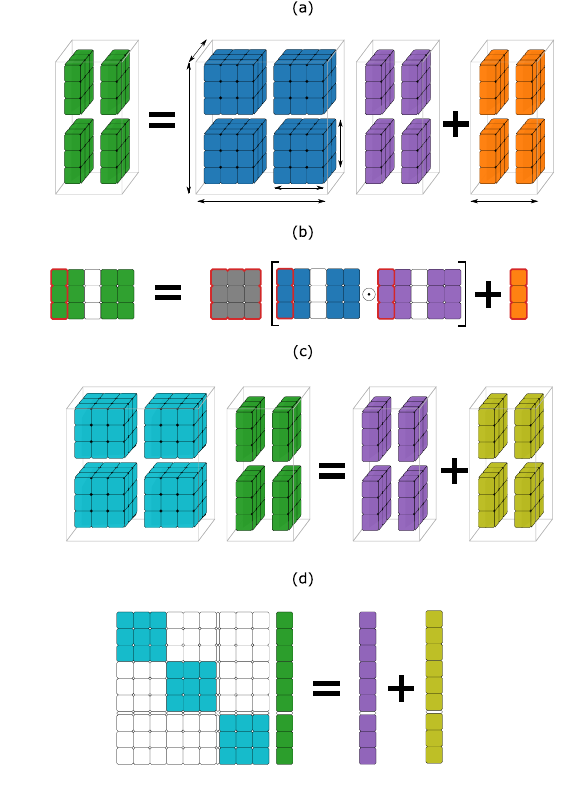
    \caption{Example of the tensor power flow formulations for a simulation with a four-dimensional power tensor $\boldsymbol{S} \in \mathbb{C}^{ p \times r \times t \times b\phi}$, for $p=r=2$ and $t=b\phi=3$. (a) Visualization of the tensor dense formulation of \eqref{eq:dense_tensor}. (b) In the case of constant power, the reshaped tensors use resources efficiently as the operations highlighted in red are performed concurrently. (c) Visualization of the tensor sparse formulation in \eqref{eq:sparse_tensor}, where tensors $\boldsymbol{\mathcal{M}}$ and $\boldsymbol{\mathcal{H}}$ are sparse. (d) Reshaped sparse formulation \eqref{eq:sparse_reshape}, which is solved iteratively using a direct sparse solver.}
    \label{fig:tensors}
    \vspace{-4mm}
\end{figure}

Consider a study that requires the analysis of an extensive number of cases of load consumption. A tensor of the power can be built as $\boldsymbol{S} \in \mathbb{C}^{\ldots \times p \times r \times t \times b\phi}$, where $p$, $r$, and $t$ could mean a number of experiments, scenarios, and time steps, with the possibility of extending the tensor to more dimensions. The \gls{fp} algorithm in \eqref{eq:fp_algorithm} in its tensor form is described as 
\begin{IEEEeqnarray}{rCl}
    \boldsymbol{V}_{(n+1)} &=& \boldsymbol{\mathcal{F}} {\boldsymbol{V}^{*}_{(n)}}^{\circ(-1)} + \boldsymbol{\mathcal{W}} \label{eq:dense_tensor}
\end{IEEEeqnarray}
where dimensions of the matrices are $\boldsymbol{\mathcal{F}} \in \mathbb{C}^{\ldots \times p \times r \times t \times b\phi \times b\phi}$, $\boldsymbol{V}_{(n+1)},\,{\boldsymbol{V}^{*}_{(n)}}^{\circ(-1)}, \boldsymbol{\mathcal{W}} \in \mathbb{C}^{\ldots \times p \times r \times b\phi \times t}$. An example of the structure of the tensors is shown in Fig.~\ref{fig:tensors}(a). Recall that the submatrices $\boldsymbol{F}$ in~\eqref{eq_b}, which form the tensor $\boldsymbol{\mathcal{F}}$, are composed by a matrix multiplication that involves $\boldsymbol{Z}_B$, meaning that $\boldsymbol{\mathcal{F}}$ is dense. It should be noted that, in the specific case of purely constant power and fixed topology, building a tensor $\boldsymbol{\mathcal{F}}$ only requires the repetition of the constant matrix $\boldsymbol{Z}_B$. Furthermore, with fixed voltage in the distribution transformer, the tensor $\boldsymbol{\mathcal{W}}$ is also constant and is the duplication of the vector $\boldsymbol{w}$ for all the cases under study. To reduce the number of operations, the repetitions of the matrix/vector, and for generalization for any number of dimensions, we define the \textit{dimensional tensor elements}, $\tau$,  as ${\tau=\ldots \times p \times r \times t}$, in order to reshape the tensors in \eqref{eq:dense_tensor} for efficient computation.

The reshaped form of the power and voltage tensors in their two-dimensional matrix form is shown in Fig.~\ref{fig:tensors}(b). Their dimensions are $\boldsymbol{\dot{S}^*},\,\boldsymbol{\dot{V}}_{(n+1)},\,{\boldsymbol{\dot{V}}^{*}_{(n)}}\in \mathbb{C}^{b\phi \times \tau}$, where the dot notation stands for the reshaped version of the tensors. The power matrix $\boldsymbol{\dot{S}^*}$ is the concatenation of the power vectors $\boldsymbol{s}^{*}$ for all the cases under study along the secondary axis. It should be mentioned that reshaping the tensor does not invalidate the convergence of the FPI algorithm, as the update of the voltage values is the same as \eqref{eq:fp_algorithm}. The advantage of the tensor form is that matrix operations can be accelerated via parallelization on the CPU using standard low-level \gls{blas}, e.g. OpenBLAS, LAPACK, IntelMKL, or exploiting the use of the multicore architectures from \glspl{gpu}, which are specifically optimized for the parallel processing matrix multiplications. Algorithm \ref{al:dense} shows the implementation of the re-shaped version of \eqref{eq:dense_tensor}. 

\begin{algorithm}[t]
\scriptsize
    \caption{: Tensor power flow - Dense}
    \label{al:dense}
    \begin{algorithmic}[1]
        \small
        \State $b\phi$ = number of bus-phases, $\tau$ = number of power flows.
        \State Input parameters: $\boldsymbol{\dot{S}} \in \mathbb{C}^{b\phi \times \tau}$, $\boldsymbol{Z}_{B} \in \mathbb{C}^{b\phi \times b\phi}$, $\boldsymbol{W} \in \mathbb{C}^{b\phi \times 1}$, tolerance, iterations.
        \State Output parameters: $\boldsymbol{\dot{V}}_{n} \in \mathbb{C}^{b\phi \times \tau}$, $n$.
        \State $\boldsymbol{\dot{V}}_0 = \mathbb{1} + \boldsymbol{j}\mathbb{0}, \quad \boldsymbol{\dot{V}}_0 \in \mathbb{C}^{b\phi \times \tau}$; $n=0$; $\mathrm{tol}=\infty$
        \While{$\mathrm{tol} \geq \mathrm{tolerance}$ $\boldsymbol{\mathrm{and}}$ $n <$ iterations}
            \For{$i = 1$ to ${\tau}$}
                \State $\boldsymbol{\dot{V}}_{(n+1)}[i]= \boldsymbol{Z}_{B}(\boldsymbol{\dot{S}}[i] \odot \boldsymbol{\dot{V}}_{(n)}^{\circ(-1)}[i])^{*} + \boldsymbol{W}$
                \State \Comment{Iterate over the dimension $\tau$ of $\boldsymbol{\dot{S}}$, and $\boldsymbol{\dot{V}}_{(n)}$.}
            \EndFor
            \State $\mathrm{tol} = \mathrm{max}(\lVert \boldsymbol{s}_{(n+1)} - \boldsymbol{s}_{(n)} \rVert^2 )$; $n = n + 1$
        \EndWhile
    \State \Return ($\boldsymbol{\dot{V}}_{(n)}$, $n$)
    \end{algorithmic}
\end{algorithm}

Although the dense formulation is simple to implement, the tensor $\boldsymbol{\mathcal{F}}$ could take a considerable amount of memory because it is dense. Therefore, for cases of networks with large $b\phi$ the sparse formulation is proposed.

\subsection{Tensor Power Flow - Sparse Formulation}
The tensor algorithm in \eqref{eq:dense_tensor} can be reformulated to exploit the sparsity of $\boldsymbol{Y}_{dd}$ and $\boldsymbol{Y}_{ds}$. The sparse formulation is defined as
\begin{IEEEeqnarray}{rCl}
    \boldsymbol{\mathcal{M}}\boldsymbol{V}_{(n+1)} &=&{\boldsymbol{V}^{*}_{(n)}}^{\circ(-1)} + \boldsymbol{\mathcal{H}}. \label{eq:sparse_tensor}
\end{IEEEeqnarray}
where ${\boldsymbol{\mathcal{M}}\in \mathbb{C}^{\ldots \times p \times r \times t \times b\phi \times b\phi}}$ is the tensor containing the resulting tensor operation of $\boldsymbol{\mathcal{M}} =- \boldsymbol{\mathcal{A}}^{\circ(-1)}\boldsymbol{\mathcal{B}}$, where $\boldsymbol{\mathcal{A}},\,\boldsymbol{\mathcal{B}}\in\mathbb{C}^{\ldots \times p \times r \times t \times b\phi \times b\phi}$ and ${\boldsymbol{\mathcal{H}}\in\mathbb{C}^{\ldots \times p \times r \times b\phi \times t}}$ a tensor containing the resulting vectors $\boldsymbol{\mathcal{H}} = \boldsymbol{\mathcal{A}}^{\circ(-1)}\boldsymbol{\mathcal{C}}$, where $\boldsymbol{\mathcal{C}}\in\mathbb{C}^{\ldots \times p \times r \times b\phi \times t}$. An example of \eqref{eq:sparse_tensor} is shown in Fig.~\ref{fig:tensors}(c). The sparse formulation does not include the inverse of $\boldsymbol{Y}_{dd}$ in any form, and similarly to the dense formulation, the expression in \eqref{eq:sparse_tensor} can be re-shaped in a two-dimensional matrix form to construct a sparse linear system of the type $Ax=b$ as

\begin{IEEEeqnarray}{rCl}
   \underbrace{\vphantom{ \boldsymbol{\dot{V}}_{(n+1)} } \boldsymbol{\dot{\mathcal{M}}}}_{A} \underbrace{\boldsymbol{\dot{V}}_{(n+1)}}_{x} &=& \underbrace{{\boldsymbol{\dot{V}}^{*}_{(n)}}^{\circ(-1)} + \boldsymbol{\dot{\mathcal{H}}}}_{b}, \label{eq:sparse_reshape}
\end{IEEEeqnarray}

\noindent where the reshaped sparse matrix $\boldsymbol{\dot{\mathcal{M}}}$ is the diagonal concatenation of the submatrices of $\boldsymbol{\mathcal{M}}$. Vectors $\boldsymbol{\dot{\mathcal{V}}}_{(n+1)},\,\boldsymbol{\dot{\mathcal{V}}}_{(n)}$, and $\boldsymbol{\dot{\mathcal{H}}}$ are vertically arranged. A visual example of the form of this reshape is shown in Fig.~\ref{fig:tensors}(d). The system in \eqref{eq:sparse_reshape} can be solved iteratively by a sparse direct solver.

The sparse direct solvers have basically three steps: (i) analysis of the matrix $\boldsymbol{\dot{\mathcal{M}}}$ to reduce \textit{fill-in} (via Cholesky, LU, or QR decomposition) and symbolic factorization, (ii) numerical factorization, and (iii) solving the system \cite{davis_direct_2006}. It is critical to note that the main advantage of \eqref{eq:sparse_reshape} is that the sparse matrix $\boldsymbol{\dot{\mathcal{M}}}$ is constant and does not change during iterations. Therefore, steps (i) and (ii) are performed only once, in the first iteration, which significantly reduces the computing time for subsequent steps. The implementation of \eqref{eq:sparse_reshape} is shown in Algorithm \ref{al:sparse}. It is worth recalling that in the NR algorithm, the sparse Jacobian matrix needs to be updated, requiring the first two steps in every iteration to calculate its inverse, increasing the computational time. Notice that rearranging the formulation from \eqref{eq:dense_tensor} to \eqref{eq:sparse_tensor} does not invalidate the convergence of the algorithm, as the sequence of the voltage values, that is, $\{\boldsymbol{V}_{(n)}\}_{n=0}^{\infty}$, is still the same.

\begin{algorithm}[t]
\scriptsize
    \caption{: Tensor power flow - Sparse}
    \label{al:sparse}
    \begin{algorithmic}[1]
        \small
        \State \textbf{Input parameters:} 
        \State $\boldsymbol{\dot{S}} \in \mathbb{C}^{b\phi \times \tau}$, $\boldsymbol{Y_{dd}} \in \mathbb{C}^{b\phi \times b\phi}, \boldsymbol{Y_{ds}} \in \mathbb{C}^{b\phi \times 1}$, tolerance, iterations.  \Comment{$\boldsymbol{Y_{dd}}$ and $\boldsymbol{Y_{ds}}$ are sparse.}
        \State \textbf{Output parameters:} $\boldsymbol{\dot{V}}_{(n)} \in \mathbb{C}^{b\phi \times \tau}$, $n$.
        \small
        \State $\boldsymbol{\dot{\mathcal{M}}} = -\mathrm{diag}(\boldsymbol{\dot{S}}_{[0]}^{*(-1)})\boldsymbol{Y_{dd}}$ \Comment{[0] first entry over dimension $\tau$.}
        \State $\boldsymbol{\dot{\mathcal{H}}} = -\mathrm{diag}(\boldsymbol{\dot{S}}_{[0]}^{*(-1)})\boldsymbol{Y_{ds}}$
        \For{$i = 1$ to ${\tau}$}
                \State $\boldsymbol{m} = -\mathrm{diag}(\boldsymbol{\dot{S}}_{[i]}^{*(-1)})\boldsymbol{Y_{dd}}$
                \State $\boldsymbol{h} = -\mathrm{diag}(\boldsymbol{\dot{S}}_{[i]}^{*(-1)})\boldsymbol{Y_{ds}}$
                \State $\boldsymbol{\dot{\mathcal{M}}} =\left(\begin{array}{c|c} 
                                                          \boldsymbol{\dot{\mathcal{M}}} & \mathbb{0} \\ 
                                                          \hline \mathbb{0} &  \boldsymbol{m}
                                                          \end{array} \right)$ \Comment{Concat. sparse matrix diagonally}
                \State $\boldsymbol{\dot{\mathcal{H}}} =\left(\begin{array}{c} 
                                                                 \boldsymbol{\dot{\mathcal{H}}}  \\ 
                                                          \hline \boldsymbol{h}
                                                          \end{array} \right)$ \Comment{Concatenate sparse vectors}
            \EndFor
        \State $\boldsymbol{\dot{V}}_0 = \mathbb{1} + \boldsymbol{j}\mathbb{0}, \quad \boldsymbol{\dot{V}}_0 \in \mathbb{C}^{(b\phi \cdot \tau) \times 1}$; $n=0$; $\mathrm{tol}=\infty$
        \While{$\mathrm{tol} \geq \mathrm{tolerance}$ $\boldsymbol{\mathrm{and}}$ $n <$ iterations}
            \State $\boldsymbol{\dot{V}}_{(n+1)} = \mathrm{solve}(\boldsymbol{\dot{\mathcal{M}}},{\boldsymbol{\dot{V}}^{*}}^{\circ(-1)}_{(n)} + \boldsymbol{\dot{\mathcal{N}}})$ \Comment{Sparse system \eqref{eq:sparse_reshape}}
            \State $\mathrm{tol} = \mathrm{max}(\lVert \boldsymbol{s}_{(n+1)} - \boldsymbol{s}_{(n)} \rVert^2 )$; $n = n + 1$
        \EndWhile
    \State $\boldsymbol{V}_n = \mathrm{reshape}(\boldsymbol{\dot{V}}_n, (b\phi \times \tau))$
    \State \Return ($\boldsymbol{V}_{(n)}$, $n$)
    \end{algorithmic}
\end{algorithm}

\section{Simulation results}\label{sec:simulations}
In this section, we discuss the comparison of Algorithms \ref{al:dense} and \ref{al:sparse}, labeled Tensor (Dense) and Tensor (Sparse), respectively, against current methods, such as \gls{sam} \cite{giraldo2022fixed}, \gls{nr} with its sparse formulation in polar coordinates (NR (Sparse))~\cite{sereeter_comparison_2019}, and the backward and forward sweep method (BFS) \cite{bompard_2000}. Additionally, the tensor-dense formulation is programmed to use a \gls{gpu} to quantify computational speed improvements; this implementation is named Tensor (GPU). The implementation of Algorithms \ref{al:dense} and \ref{al:sparse} is publicly available to the community as a Python package named TensorPowerFlow\footnote{The public link to the repository will be provided after the acceptance of the paper}. The experiments were conducted on a conventional laptop with an Intel(R) Core(TM) i7-7700HQ CPU @ 2.80GHz, with 8 Logical Processor(s), and a NVIDIA Quadro M1200 with 4GB (GDDR5) of memory.
\subsection{Computational Performance}
The goal of the experiments is to test the computation performance of the different methods for two aspects: (i) grid size and (ii) dimensionality. For the first aspect, different networks with bus-phases $b\phi$, which range from 9 to 5k, are generated using a $k$-ary tree generative model, with random $k$-child between 1 and 5, in order to simulate the radial structure of the distribution system \cite{storer_2002}. For the second aspect, the dimensional tensor elements $\tau$, which are the number of PFs to compute, range from 10 to 525k, and the power values are generated using a multivariate elliptical copula~\cite{duque_conditional_2021} to simulate realistic scenarios of consumption profiles. The same power scenarios and networks are used for all methods. Additionally, for methods that use sparse solvers, the Intel oneMKL - PARDISO is used in all of them for a fair comparison.

\begin{figure}[t]
\centering
\vspace{-3mm}
\includegraphics[width=\columnwidth]{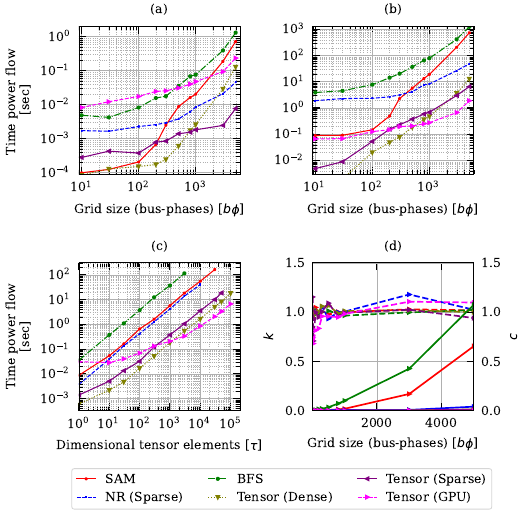}
\vspace{-8mm}
\caption{Comparison of the performance of the algorithms. Computational wall-times ($t_c$) increase the bus-phases size, $b\phi$, for (a) 1 power flow, and (b) 500 power flows. (c)~Test for increased dimensional tensor elements $\tau$, for a grid size of 500 bus-phases ($b\phi$). (d) Computational complexity fit with asymptotic complexity model $t_c = c \cdot n^k$ (solid lines are $c$, dotted $k$). Tensor \textit{dense}, \textit{sparse}, and \textit{GPU} are the proposed algorithms in this paper. }
\label{fig:performance}
\vspace{-5mm}
\end{figure}

The performance of the methods for the increase of $b\phi$ is shown for one PF in Fig.~\ref{fig:performance}(a), and for 500 PF Fig.~\ref{fig:performance}(b). The best methods are SAM and Tensor (Dense) for one PF with an execution time of 0.1 [ms] for the smallest network. However, both methods reduce its performance for an increased grid size, where around 900 bus-phases, the preference is for the Tensor (Sparse) formulation. In the case of 500 PF (Fig.\ref{fig:performance}(b)), the SAM lags behind because it requires more matrix multiplications in its original formulation; the Tensor (Dense) continues as the top performer for smaller grids, seconded by the Tensor (Sparse), where after a grid size of 2k is better than the Tensor (Dense). An interesting behavior is seen for the Tensor (GPU) implementation, which is significantly lower for a small amount of power flows (Fig.~\ref{fig:performance}(a)), which is explained due to the overhead time required to transfer the data from the host computer to the \gls{gpu}. However, when the number of matrix operations is large, either for increase $\tau$ or $b\phi$, the Tensor (GPU) is the best performer, as shown in Fig.~\ref{fig:performance}(b)(c). The computational complexity of the algorithms with a model of the form $t_c=c\,n^k$, is shown Fig.~\ref{fig:performance}(d), confirming that SAM and BFS scale poorly for larger grids.

\begin{table}[t]
\centering
\caption{Computing Wall-times for solving one year of power flows at different time resolutions (table values in minutes).}\label{table:exp_results}
\scalebox{0.85}{
\begin{tabular}{cccccc}
\toprule
Algorithm &
  \begin{tabular}[c]{@{}c@{}}Grid size \\ ($b\phi$)\end{tabular} &
  \begin{tabular}[c]{@{}c@{}}@1 hr \\ (8760 pf)\end{tabular} &
  \begin{tabular}[c]{@{}c@{}}@30 min \\ (17520 pf)\end{tabular} &
  \begin{tabular}[c]{@{}c@{}}@15 min\\ (35040 pf)\end{tabular} &
  \begin{tabular}[c]{@{}c@{}}@1 min\\ (525600 pf)\end{tabular} \\ \midrule
BFS             & 100  & 1.98   & 3.96   & 7.91   & 118.71   \\
NR (Sparse)     & 100  & 1.03   & 2.06   & 4.12   & 61.85    \\
Tensor (Sparse) & 100  & 0.17   & 0.34   & 0.68   & 10.14    \\
SAM             & 100  & 0.02   & 0.04   & 0.09   & 1.29     \\
Tensor (GPU)    & 100  & 0.02   & 0.04   & 0.09   & 1.30     \\
Tensor (Dense)  & 100  & \textbf{0.01}   & \textbf{0.01}   & \textbf{0.02}   & \textbf{0.37}     \\ \midrule
BFS             & 5000 & 186.49 & 372.99 & 745.97 & - \\
NR (Sparse)     & 5000 & 8.77   & 17.54  & 35.08  & 526.23   \\
Tensor (Sparse) & 5000 & 2.43   & 4.86   & 9.71   & 145.71   \\
SAM             & 5000 & 108.09 & 216.18 & 432.36 & -  \\
Tensor (GPU)    & 5000 & \textbf{0.92}   & \textbf{1.84}   & \textbf{3.69}   & \textbf{55.34}    \\
Tensor (Dense)  & 5000 & 2.60   & 5.19   & 10.38  & 155.71  \\
\bottomrule
\end{tabular}}
\vspace{-4mm}
\end{table}

\subsection{Yearly Time Series Simulation}
A comparative test is performed for a yearly \gls{tss} study for grid sizes ranging from 100 to 5000 bus-phases~($b\phi$), for time resolution of 1, 15, 30, and 60 minutes. The results are shown in Table~\ref{table:exp_results}, where the maximum waiting time for the results is set to 9 hours. In general, all the \gls{fp}  methods are preferred and best suited for large amounts of \glspl{pf} instead of conventional NR and BFS techniques. Among the iterative methods, the proposed Tensor (Dense) algorithm stood out as the fastest for the smaller grid size case. It completed the 1-minute resolution case (approximately 525k \glspl{pf} calculations) in only 22 seconds. To put this in perspective, when compared with the NR (Sparse) method --which took more than one hour to accomplish the same task-- the Tensor (Dense) revealed a remarkable speedup factor of 164. In the case of larger grids (5k bus-phases), the proposed methods are still more efficient than NR (Sparse), and the Tensor (Sparse) is the second best for all time resolutions. However, the \gls{gpu} implementation of Tensor (Dense) is the fastest, showing the speed up from \gls{gpu}, which is capable of massively parallelizing the matrix multiplications automatically in a multi-thread setting. It is worth noting that a GPU-based sparse solver could be utilized to enhance the Tensor (Sparse), though this area requires further exploration.



\section{Conclusions}\label{sec:conclusions}
This paper presented practical \gls{pf} formulations for analyses of distribution systems in a multidimensional scope. 
The formulations are based on a \gls{fp}  algorithm, and two algorithms are presented in their tensor forms for dense and sparse versions. The convergence proof of the algorithms and the existence of a solution are presented with its geometrical and physical interpretation. The mathematical analysis shows that the \gls{fp} algorithm converges to the high impedance value in the case of the existence of a solution, and it can be extended to the multidimensional formulation. The performance of the proposed algorithms compared to conventional methods such as NR and BFS was evaluated in two aspects: the size of the grid (number of bus-phases, $b\phi$) and dimensionality (number of PFs, $\tau$). 
The algorithms were tested in a practical  \gls{tss} application for different grid sizes and time resolutions. Tensor~(Dense) emerged as the fastest algorithm, providing a significant speedup over traditional methods like NR (Sparse), especially for smaller grid sizes. The study also introduced a GPU implementation, Tensor (GPU), which showed reduced performance for a small number of power flows due to data transfer overhead but excelled when the number of matrix operations was substantial, as seen when the grid size and dimensionality increased.



\bibliographystyle{IEEEtran}
\bibliography{bibliography}

\begin{thebibliography}{10}
\providecommand{\url}[1]{#1}
\csname url@samestyle\endcsname
\providecommand{\newblock}{\relax}
\providecommand{\bibinfo}[2]{#2}
\providecommand{\BIBentrySTDinterwordspacing}{\spaceskip=0pt\relax}
\providecommand{\BIBentryALTinterwordstretchfactor}{4}
\providecommand{\BIBentryALTinterwordspacing}{\spaceskip=\fontdimen2\font plus
\BIBentryALTinterwordstretchfactor\fontdimen3\font minus
  \fontdimen4\font\relax}
\providecommand{\BIBforeignlanguage}[2]{{%
\expandafter\ifx\csname l@#1\endcsname\relax
\typeout{** WARNING: IEEEtran.bst: No hyphenation pattern has been}%
\typeout{** loaded for the language `#1'. Using the pattern for}%
\typeout{** the default language instead.}%
\else
\language=\csname l@#1\endcsname
\fi
#2}}
\providecommand{\BIBdecl}{\relax}
\BIBdecl

\bibitem{Qureshi_2019}
M.~U. Qureshi, S.~Grijalva, M.~J. Reno, J.~Deboever, X.~Zhang, and R.~J.
  Broderick, ``A fast scalable quasi-static time series analysis method for
  {PV} impact studies using linear sensitivity model,'' \emph{IEEE Trans. Sust.
  Energy}, vol.~10, no.~1, pp. 301--310, 2019.

\bibitem{liu2018probabilistic}
C.~Liu, K.~Sun, B.~Wang, and W.~Ju, ``Probabilistic power flow analysis using
  multidimensional holomorphic embedding and generalized cumulants,''
  \emph{IEEE Trans. Power Syst.}, vol.~33, no.~6, pp. 7132--7142, 2018.

\bibitem{duque2022community}
E.~M.~S. Duque, J.~S. Giraldo, P.~P. Vergara, P.~Nguyen, A.~van~der Molen, and
  H.~Slootweg, ``Community energy storage operation via reinforcement learning
  with eligibility traces,'' \emph{Electr. Power Syst. Res.}, vol. 212, p.
  108515, 2022.

\bibitem{Garces_2019}
A.~Garcés, J.~J. Mora, and M.-A. Useche, ``Putting tensors back in power
  systems analysis,'' in \emph{2019 Int. Conf. on Smart Energy Systems and
  Technologies (SEST)}, 2019, pp. 1--5.

\bibitem{nickolls_gpu_2010}
J.~Nickolls and W.~J. Dally, ``The {GPU} {Computing} {Era},'' \emph{IEEE
  Micro}, vol.~30, no.~2, pp. 56--69, Mar. 2010.

\bibitem{zhou_gpu_2017}
G.~Zhou, R.~Bo, L.~Chien, X.~Zhang, F.~Shi, C.~Xu, and Y.~Feng,
  ``\BIBforeignlanguage{en}{{GPU}-{Based} {Batch} {LU}-{Factorization} {Solver}
  for {Concurrent} {Analysis} of {Massive} {Power} {Flows}},''
  \emph{\BIBforeignlanguage{en}{IEEE Trans. Power Syst.}}, vol.~32, no.~6, pp.
  4975--4977, Nov. 2017.

\bibitem{li_gpu_2017}
X.~Li, F.~Li, H.~Yuan, H.~Cui, and Q.~Hu,
  ``\BIBforeignlanguage{en}{{GPU}-{Based} {Fast} {Decoupled} {Power} {Flow}
  {With} {Preconditioned} {Iterative} {Solver} and {Inexact} {Newton}
  {Method}},'' \emph{\BIBforeignlanguage{en}{IEEE Trans. Power Syst.}},
  vol.~32, no.~4, pp. 2695--2703, Jul. 2017.

\bibitem{zhou_gpu_2018}
G.~Zhou, R.~Bo, L.~Chien, X.~Zhang, S.~Yang, and D.~Su,
  ``\BIBforeignlanguage{en}{{GPU}-{Accelerated} {Algorithm} for {Online}
  {Probabilistic} {Power} {Flow}},'' \emph{\BIBforeignlanguage{en}{IEEE Trans.
  Power Syst.}}, vol.~33, no.~1, pp. 1132--1135, Jan. 2018.

\bibitem{zhou_gpu_acc_2017}
G.~Zhou, Y.~Feng, R.~Bo, L.~Chien, X.~Zhang, Y.~Lang, Y.~Jia, and Z.~Chen,
  ``\BIBforeignlanguage{en}{{GPU}-{Accelerated} {Batch}-{ACPF} {Solution} for
  {N}-1 {Static} {Security} {Analysis}},'' \emph{\BIBforeignlanguage{en}{IEEE
  Trans. Smart Grid}}, vol.~8, no.~3, pp. 1406--1416, May 2017.

\bibitem{giraldo2022fixed}
J.~S. Giraldo, O.~D. Montoya, P.~P. Vergara, and F.~Milano, ``A fixed-point
  current injection power flow for electric distribution systems using
  {Laurent} series,'' \emph{Electr. Power Syst. Res.}, vol. 211, p. 108326,
  2022.

\bibitem{grainger_power_1994}
J.~J. Grainger and W.~D. Stevenson, \emph{\BIBforeignlanguage{eng}{Power system
  analysis}}, ser. {McGraw}-{Hill} series in electrical and computer
  engineering {Power} and energy.\hskip 1em plus 0.5em minus 0.4em\relax New
  York, NY St. Louis San Francisco: McGraw-Hill, Inc, 1994.

\bibitem{sur_existence_2019}
U.~Sur and G.~Sarkar, ``\BIBforeignlanguage{en}{Existence of {Explicit} and
  {Unique} {Necessary} {Conditions} for {Power} {Flow} {Insolvability} in
  {Power} {Distribution} {Systems}},'' \emph{\BIBforeignlanguage{en}{IEEE Syst.
  Jour.}}, vol.~13, no.~1, pp. 702--709, Mar. 2019.

\bibitem{bolognani_existence_2016}
S.~Bolognani and S.~Zampieri, ``\BIBforeignlanguage{en}{On the {Existence} and
  {Linear} {Approximation} of the {Power} {Flow} {Solution} in {Power}
  {Distribution} {Networks}},'' \emph{\BIBforeignlanguage{en}{IEEE Trans. Power
  Syst.}}, vol.~31, no.~1, pp. 163--172, Jan. 2016.

\bibitem{yu_simple_2015}
S.~Yu, H.~D. Nguyen, and K.~S. Turitsyn, ``\BIBforeignlanguage{en}{Simple
  certificate of solvability of power flow equations for distribution
  systems},'' in \emph{\BIBforeignlanguage{en}{2015 {IEEE} {Power} \& {Energy}
  {Society} {General} {Meeting}}}.\hskip 1em plus 0.5em minus 0.4em\relax
  Denver, CO, USA: IEEE, Jul. 2015, pp. 1--5.

\bibitem{ahmadi_fast_2022}
A.~Ahmadi, M.~C. Smith, E.~R. Collins, V.~Dargahi, and S.~Jin,
  ``\BIBforeignlanguage{en}{Fast {Newton}-{Raphson} {Power} {Flow} {Analysis}
  {Based} on {Sparse} {Techniques} and {Parallel} {Processing}},''
  \emph{\BIBforeignlanguage{en}{IEEE Trans. Power Syst.}}, vol.~37, no.~3, pp.
  1695--1705, May 2022.

\bibitem{davis_direct_2006}
T.~A. Davis, \emph{Direct methods for sparse linear systems}, ser. Fundamentals
  of algorithms.\hskip 1em plus 0.5em minus 0.4em\relax Philadelphia: SIAM,
  2006.

\bibitem{sereeter_comparison_2019}
B.~Sereeter, C.~Vuik, and C.~Witteveen, ``\BIBforeignlanguage{en}{On a
  comparison of {Newton}–{Raphson} solvers for power flow problems},''
  \emph{\BIBforeignlanguage{en}{J. Comput. Appl. Math.}}, vol. 360, pp.
  157--169, Nov. 2019.

\bibitem{bompard_2000}
E.~Bompard, E.~Carpaneto, G.~Chicco, and R.~Napoli,
  ``\BIBforeignlanguage{en}{Convergence of the backward/forward sweep method
  for the load-flow analysis of radial distribution systems},''
  \emph{\BIBforeignlanguage{en}{Int. J. Electr. Power Energy Syst.}}, vol.~22,
  no.~7, pp. 521--530, Oct. 2000.

\bibitem{storer_2002}
J.~A. Storer, \emph{\BIBforeignlanguage{en}{An {Introduction} to {Data}
  {Structures} and {Algorithms}}}.\hskip 1em plus 0.5em minus 0.4em\relax
  Boston, MA: Birkhäuser Boston, 2002.

\bibitem{duque_conditional_2021}
E.~M.~S. Duque, P.~P. Vergara, P.~H. Nguyen, A.~Van Der~Molen, and J.~G.
  Slootweg, ``Conditional {Multivariate} {Elliptical} {Copulas} to {Model}
  {Residential} {Load} {Profiles} {From} {Smart} {Meter} {Data},'' \emph{IEEE
  Trans. Smart Grid}, vol.~12, no.~5, pp. 4280--4294, Sep. 2021.

\end{thebibliography}

\end{document}